\documentclass[twocolumn,prX]{revtex4}%

\usepackage{amsmath}%
\usepackage{amsfonts}%
\usepackage{amssymb}%
\usepackage{graphicx}
\newcommand{\ket}[1]{|#1\rangle}
\newcommand{\bra}[1]{\langle #1|}
\newcommand{\braket}[1]{\langle #1 \rangle}
\DeclareMathOperator{\tr}{tr}
\newcommand{\aup}{\hat{a}^{\dag}}
\newcommand{\adown}{\hat{a}}
\usepackage{graphics}
\begin{document}

\title{Cooling a micro-mechanical resonator to its ground state by measurement back-action}
\author{Christian Bergenfeldt$^{\dag}$ and Klaus M\o lmer$^{\ddag}$\thanks{This is for making an acknowledgement.}
\\$^{\dag}$Department of Physics, Lund University, Box 118, 22100 Lund, Sweden
\\$^{\ddag}$Lundbeck Foundation Center for Theoretical Quantum System Research, Department of Physics and Astronomy\\ Aarhus University, DK 8000 Aarhus C, Denmark}
\date{June 10, 2009}

\begin{abstract}
We present an analysis of the cooling of a micro-mechanical resonator by means of measurements and back action. The measurements are performed via the coupling to a Cooper-pair box, and although the coupling does not lead to net cooling, the extraction of information and hence entropy from the system leads to a pure quantum state. Under suitable circumstances, the states become very close to coherent states, conditioned on the measurement record, and can hence be displaced to the oscillator ground state.
\end{abstract}

\maketitle

\section{Introduction}

Micro-mechanical resonators are elastic micron-sized beams fixed in one or both ends with tranversal modes of vibration with frequencies in the range 10 to 1000 Mhz \cite{armour2002ead}. Applications of these oscillators include ultrasensitive force detectors, electrometers and displacement detectors. Recently, interest has been directed towards developing schemes to make these resonators display non-classical behaviour. Here, one tries to cool the resonators to their vibrational ground-state \cite{jaehne2008gsc, hopkins2003fcn, zippilli2008ccn} and to generate different non-classical states of motion \cite{armour2002ead,irish2003qmc}.  The quantum regime represents the fundamental resolution limit for these systems when they are used for ultra-sensitive detection purposes, and by making special, squeezed states or macroscopic superposition states, one may significantly enhance their performance. When operated in the quantum regime, mechanical resonators may find applications as quantum information storage devices and as mediators of quantum states between other quantum systems. Also, proposals exist to produce quantum correlated states of motion of two oscillators
\cite{hartmann2007sse} and to establish quantum entangled states of hybrid systems involving a mechanical resonator and, e.g., a quantized field \cite{vitali2007enr} or an atomic cloud \cite{hammerer2009eep}.

For all of these ideas to be realized, however, it is necessary to coherently control the quantum state of these systems, and with their low vibrational frequencies, even at cryogenic temperatures, their thermal excitations are significant, and the quantum state will in general be a thermal mixture. Different means of cooling have thus been proposed and in particular mechanical resonators which also act as optical resonators, and which can be detuned with respect to the frequency of incident laser radiation, have been cooled in analogy with frictional and sideband laser cooling of atoms and ions
\cite{schliesser97rpc,schliesser2008rsc,groeblacher2009dum} .
Other means for cooling, e.g., by the back action of the coupling to resonant circuit elements \cite{naik2006cnr} have been analyzed and demonstrated, and in this article we present a way to turn a mixed thermal state into a pure coherent state of motion using the quantum mechanical back-action due to measurements on the mechanical oscillator. We imagine these measurements to take place by a Cooper-pair box which is capacitively coupled to the resonator and which is read out by a single electron transistor (SET). The measurement does not cool the system, but it extracts entropy and provides a pure coherent state, and hereafter, or along with the meaurement process, a simple feedback mechanism may be used to displace the state, so that the mechanical ground state is eventually obtained. In our simulations we are able to estimate the efficiency of this method and to incorporate various effects such as finite interaction time and coupling to a thermal reservoir, and see how they affect the cooling procedure. In Sec. II, we introduce our physical system. In Sec. III, we describe the measurement protocol and the formalism describing the effect of measurements on the oscillator. In Sec. IV, we present our simulation results. In Sec. V we describe the use of feedback to reach the oscillator ground state, and we show how well the measurements and feedback are able to balance the heating of the oscillator due to its coupling to a thermal reservoir. Sec. VI concludes the paper.

\section{The system}

The micro-mechanical resonator studied is a cantilever beam. This is a beam rigidly attached in one end and completely free in the other end. The beam has flexural modes of vibration in the directions perpendicular to the axis of the beam. The lowest frequency mode can be modelled  as a harmonic oscillator and one can assume weak coupling between this mode and the other modes of the cantilever \cite{armour2002ead}.

The other part of the system is a Cooper-pair box. This is, effectively, a charged capacitor with an anharmonic ladder of discrete charge states, and where a two-level system can be effectively identified with two specific charge states, $\ket{-}=\ket{n}$ and $\ket{+}=\ket{n+1}$. These states correspond to $n$ or $n+1$ Cooper-pairs being present on the capacitor "island" of the Cooper-pair box, respectively. When restricted to these two states, the Hamiltonian of the Cooper-pair box can be written \cite{bouchiat1998qcs}:
\begin{equation}
H_{CPB}=E_{c}\delta n\sigma_{z}-\frac{E_{J}}{2}\sigma_{x}
\label{eq:hamiltoniancpb}
\end{equation}
Here $E_{C} $ is the so-called charging energy of the Cooper-pair box, $\delta n$ is a tunable dimensionless parameter taking values between -1/2 and 1/2 and $E_{J}$ is the Josephson energy. The operators $\sigma_{x},\sigma_{y}$ and $\sigma_{z}$ are defined by their action on the states $\ket{-}=\ket{n}$ and $\ket{+}=\ket{n+1}$ the same way as the corresponding Pauli operators acts on the eigenstates of the projection of a spin onto the z-axis. If Cooper-pair box parameters are chosen \cite{armour2002ead} such that the second term in the Hamiltonian can be neglected, the charge states are eigenstates of the system.

\begin{figure}
	\begin{center}
		\resizebox{!}{50mm}{\includegraphics{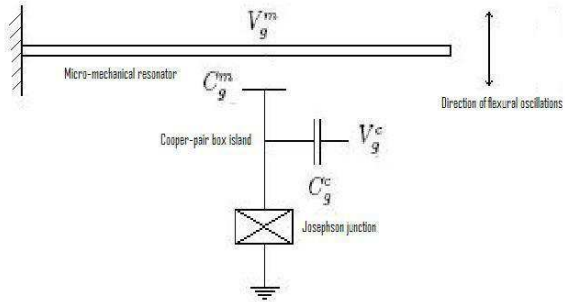}}
	\end{center}
	\caption{This figure shows a sketch of the setup. The double arrow shows the direction in which the oscillations are imagined to take place. $V_{g}^{m}$ and $V_{g}^{c}$ are the voltages applied to the micro-mechanical oscillator and the other gate electrode, respectively. $C_{g}^{m}$ and $C_{g}^{c}$ are capacitances between the electrodes and the Cooper-pair box "island".}
	\label{setup}
\end{figure}

In Fig. \ref{setup} a schematic sketch of the setup is shown. We see that by putting an electric voltage on the cantilever we will get a capacitive coupling between the systems. The coupling between the Cooper-pair box and all other modes but one of the two lowest frequency modes will be weak and can thus be neglected \cite{armour2001ewi}. The coupling between all modes is also weak. Our system therefore consists of a Cooper-pair box coupled to a harmonic oscillator. When $E_J=0$, the Hamiltonian of the combined system is \cite{armour2002ead}:
\begin{equation}
H=\hbar\omega\hat{a}^{\dag}\hat{a}+E_{c}\delta
n\sigma_{z}+\lambda(\hat{a}^{\dag}+\hat{a})\sigma_{z}
\label{eq:hamiltonian}
\end{equation}
where $\omega$ is the frequency of the oscillator and where $\lambda$ is the coupling strength between the systems, depending on the displacements of the cantilever by $x\propto \hat{a}+\hat{a}^\dagger$ from its equillibrium position and being proportional to the voltage between the systems \cite{armour2002ead}.

The Cooper-pair box has two advantageous properties. First, $E_{J}$ can be switched to values much larger than $E_{c}\delta n$ for a short period of time $\Delta t$ \cite{shnirman1997qms}. The box is then said to be taken to its degeneracy point and coherent transitions are driven between the charge states. For $\frac{E_{J}\Delta t}{2\hbar}$ equal to the angle $\delta$, we say that we have applied a $\delta$-pulse  causing a Rabi oscillation between the charge states of the Cooper-pair box. This is a unitary process, and applying the $\delta$-pulse to the Cooper-pair box in the arbitrary initial superposition state $\ket{\psi}=c_{1}\ket{-}+c_{2}\ket{+}$  we transfer the system to the new superposition state $(c_{1}\cos(\delta)+c_{2}i\sin(\delta))\ket{-}+(c_{2}\cos(\delta)+c_{1}i\sin(\delta))\ket{+}$. By switching on an off $E_{J}$  for short intervals of time, and by letting the Cooper-pair box and cantilever oscillator interact, the two systems become entangled. Second, using a radio-frequency single electron transistor (rf-SET), it is possible with high sensitivity to measure the charge state of the Cooper-pair box \cite{armour2002ead}. Currently, with realistic Cooper-pair box parameters it is possible to measure the charge of the Cooper-pair box within a time $\tau_{m}=4 ns$ \cite{aassime2001rfs}. This is much shorter than the cantilever time scale of motion, and we will for simplicity of our analysis assume that measurements are instantaneous with respect to the dynamics  of the oscillator. Recent experiments have demonstrated an alternative read out of a Cooper-pair box by its modification of the transmission properties of a stripline cavity \cite{readout1}.
By carrying out measurements of the charge states of the Cooper pair box, we can extract information, and thus remove uncertainty about the cantilever motion.

\section{Measurements}

We want to gain information about the state of the oscillator, and we propose the following measurement procedure. At first the Cooper-pair box is in either one of the charge states. Then a fast $\pi/2$-pulse is applied to the Cooper-pair box transferring it almost instantaneously into an even super-position of the charge states. After a time interval $\tau$ during which the Cooper-pair box and the cantilever interact, a second $\pi/2$-pulse is applied to the Cooper-pair box, and then the charge of the Cooper-pair box is measured, returning it to one of the charge states, so that the procedure can be repeated over an over again.

We can give the following picture of how the measurement procedure provides the information about the state of the oscillator: After the first $\pi/2$-pulse puts the Cooper-pair box in a super-position of the charge states, the interaction Hamiltonian terms containing $\sigma_{z}$ will build up a relative phase $\Delta \phi$ between the two charge states. Since one of the terms containing $\sigma_{z}$ in the Hamiltonian \eqref{eq:hamiltonian} depends on $x\propto(\aup+\adown)$ the phase $\Delta \phi$ will depend on the state of the oscillator mode. The second $\pi/2$-pulse will rotate the state of the Cooper-pair box so that the relative phase $\Delta \phi$ is converted into a population difference, and hence, a measurement of the charge state populations yields the information about the  oscillator state.
To properly account for this information retriveal and its consequences for the dynamics of the cantilever quantum state conditioned on the read-out sequence, we describe the dynamics and the measurements in a density operator language. Let $\rho$ be the density operator of the motional oscillator state prior to a measurement, and assume that the Cooper-pair box is in the $\ket{-}$ state (the treatment of the Cooper-pair box being initially in the $\ket{+}$ state is fully equivalent). After the  first $\pi/2$-pulse we have for the combined system the density operator:
\begin{equation}
\tilde{\rho}=\frac{\rho\otimes (\ket{-}\bra{-}+i\ket{+}\bra{-}-i\ket{-}\bra{+}+\ket{+}\bra{+})}{2}
	\label{eq:pi2}
\end{equation}
After an interaction time $\tau$ the density operator becomes $U\tilde{\rho} U^{\dag}$, where $U=\exp(\frac{-iH\tau}{\hbar})$ with the Hamiltonian in \eqref{eq:hamiltonian}. Since the charge states are eigenstates of \eqref{eq:hamiltonian} we can get a simple expression for $U\rho U^{\dag}$ by introducing the operators $U_{\pm}=\exp(-i\omega\tau(\hat{a}^{\dag}\hat{a}\pm(\frac{E_{c}\delta n}{\hbar\omega}+\kappa(\hat{a}^{\dag}+\hat{a}))))$, acting only on the oscillator space. Using this notation, we get $U\big( \rho\otimes(\ket{\alpha}\bra{\beta})\big)U^{\dag}=U_{\alpha}\rho U_{\beta}^{\dag}\otimes(\ket{\alpha}\bra{\beta})$, with $\alpha,\beta=+$ and $-$, and $U\tilde{\rho} U^{\dag}$ can hence be written

\begin{align}
U \tilde{\rho} U^{\dag} = \nonumber \\
\frac{1}{2}( \rho_{--}\otimes (\ket{-}\bra{-})+i\rho_{+-} \otimes (\ket{+}\bra{-}) \nonumber \\
-i\rho_{-+}\otimes (\ket{-}\bra{+})+\rho_{++}\otimes (\ket{+}\bra{+}))
\end{align}

with $\rho_{\alpha\beta}=U_{\alpha}\rho U_{\beta}^{\dag}$. After the final $\pi/2$-pulse on the Cooper-pair box we have the state:
\begin{align}
\frac{1}{4}( (\rho_{--}-\rho_{-+}-\rho_{+-}+\rho_{++})\otimes (\ket{-}\bra{-})\nonumber\\
-i(\rho_{--}+\rho_{-+}-\rho_{+-}-\rho_{++})\otimes (\ket{-}\bra{+})\nonumber\\
+i(\rho_{--}+\rho_{-+}-\rho_{+-}-\rho_{++})\otimes (\ket{+}\bra{-})\nonumber\\
(\rho_{--}+\rho_{-+}+\rho_{+-}+\rho_{++})\otimes (\ket{+}\bra{+}) )
	\label{eq:}
\end{align}

At this stage the projective measurement is performed on the Cooper-pair box charge state basis, i.e., the state is replaced by the first (last) line of this expression if the Cooper-pair box is found in state $\ket{-(+)}$. Introducing operators $M_{\mp}=\frac{U_{-}\mp U_{+}}{2}$ the effective modification of the cantilever state due to the interaction with the Cooper-pair box and its subsequent detection can therefore be written
$\rho \rightarrow  \rho_{\mp}=\frac{M_{\mp}\rho M_{\mp}^{\dag}}{\tr (M_{\mp}^{\dag}M_{\mp}\rho)}$, occurring with the probabilities $P_{\mp}=\tr(M_{\mp}^{\dag}M_{\mp}\rho)$.  The operators $U_\pm$ are exponential in $\hat{a}, \hat{a}^\dagger$, and $\hat{a}^\dagger\hat{a}$, i.e., they are analytically known combinations of a displacement and a rotation in the oscillator phase space and can be readily evaluated, e.g., in the eigenstate basis of the oscillator. This dynamics can thus be readily simulated on a computer.

\section{Simulations}

Having derived the measurement operators $M_{\mp}$ we can numerically simulate the measurement procedure using random numbers to decide between the measurement outputs in each step of the measurement sequence described above.

Initially, the system is assumed to be in thermal equillibrium with a heat reservoir of temperature $T$. The energy difference between the Cooper-pair box charge states is assumed to be large compared to $kT$. Hence, we can assume that it is initially in the state $ \ket{-}$ and the initial density operator can therefore be written $\tilde{\rho}=\rho\otimes(\ket{-}\bra{-})$ with :
\begin{equation}
\rho=\sum_{n}\frac{\exp(\hbar\omega/kT)-1}{\exp(\hbar\omega/kT)}\exp(-\hbar\omega n/kT)\ket{n}\bra{n} 	
	\label{eq:thermal}
\end{equation}
In the following we take an oscillator frequency of $\omega=50$MHz and an initial temperature of 50 mK, so that the oscillator occupies the lowest hundred oscillator states. In our simulations we assume an interaction time $\tau=0.1/\omega$ and the charge state energy separation is given by $E_{c}\delta n=100\hbar\omega$. In our simulations we determine the conditioned dynamics of the cantilever over 160 subsequent interaction and measurement sequences, and we repeat such simulations 200 times to be able to characterize the typical properties of our detection records. Note, however, that we do not form an ensemble averaged density operator by simply adding together the individual simulation records: the experiment is supposed to take place on a single cantilever, and the specific detection record of a single run is assumed to be available to the experimentalist, who can thus infer the cantilever quantum state according to the above analysis.

\begin{figure}[htb]
	\begin{center}
		\resizebox{!}{50mm}{\includegraphics{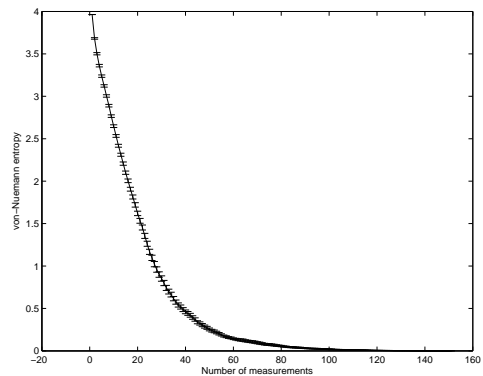}}
	\end{center}
	\caption{The averaged von-Neumann entropy of the cantilever state conditioned on the measurement record is plotted against the number of measurements performed in over 200 independent simulations. Parameters used are $\omega=50$MHz, $\lambda=1.5\hbar\omega$, $E_{c}\delta n=100\hbar\omega$ and an initial temperature of $T=50$mK.}
	\label{entropy}
\end{figure}

\begin{figure}[htb]
	\begin{center}
		\resizebox{!}{70mm}{\includegraphics{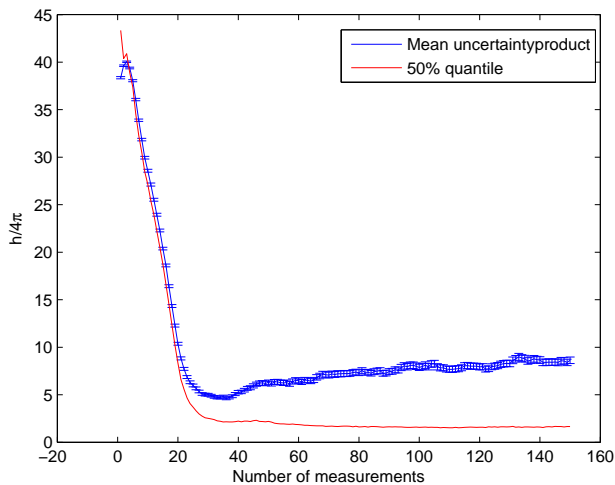}}
	\end{center}
	\caption{(Color online) The upper curve shows the mean value of the position-momentum uncertainty product as a function of time obtained from sampling 2000 independent iterations. 50 \% of the realizations have their uncertainty product below the lower curve in the figure (the 50 \% quantile). The same parameters were used for the simulations as in Fig. \ref{entropy}}
	\label{uncertainty}
\end{figure}

We first address the approach of our cantilever towards a pure quantum state as a consequence of more and more information about the state becoming known via the measurements. For this purpose we calculate in each run the von-Neumann entropy of the time dependent density operator, and in Fig. \ref{entropy} we plot the average over all realizations of this quantity as function of time. As is clearly seen, the mean value of the von-Neumann entropy decreases with the number of measurements performed and tends to zero after sufficiently many measurements. After a hundred measurements the entropy is practically zero and we have full knowledge of the pure state of the oscillator in every realization.
\\
\\
Having thus asserted the successful convergence of the quantum state towards a pure state, i.e., a wave function, we will now identify what state is actually produced. It is natural to quantify the state by the position and momentum variables, and in Fig. \ref{uncertainty} we show the uncertainty product of these observables, averaged over 2000 simulations with the same parameters that were used in the previous calculations. We see that the position-momentum uncertainty product decreases and flattens out after about 50 measurements. It does not reach the minimum value allowed by the Heisenberg uncertainty relation but is only about a factor 5-10 times larger than this minimum. Examining a little closer the individual realizations, we find variations, and plotting the 50 percent quantile, i.e., the value  which is above the uncertainty product of the 50 percent of the realizations with the smallest uncertainty products, we observe results close to the Heisenberg limit. In these cases we are very close to minimum uncertainty states, and we know that they are very close to either coherent states or squeezed states of motion.
With the assumption that the majority of realizations converge to coherent states, we make a simple coherent state fidelity analysis by computing after 150 measurements for every realization the state overlap $\braket{\alpha|\rho|\alpha}$ between the conditioned state $\rho$ and a coherent state of motion $\ket{\alpha}$, having the same mean position and momentum.
\begin{figure}[htb]
	\begin{center}
		\resizebox{!}{50mm}{\includegraphics{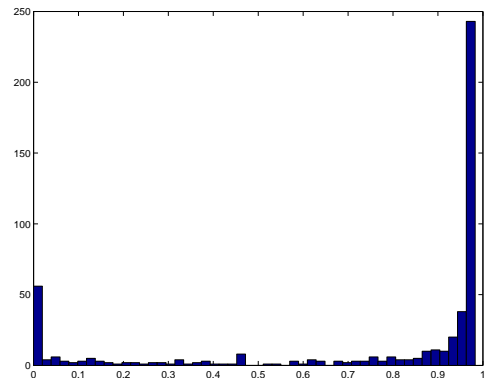}}
	\end{center}
	\caption{(Color online) Histogram of values of the coherent state fidelity after 150 measurements for 500 simulations.}
	\label{newfid}
\end{figure}
A histogram of these results is shown in Fig. \ref{newfid} for 500 runs. As can be seen a large fraction of the runs results in states which have a coherent state fidelity above 0.9, but we also find a sizable fraction with a nearly vanishing coherent state overlap. To explain these findings, in Fig. \ref{Qcanti} the Q-function, $Q(\alpha)=\langle\alpha|\rho|\alpha\rangle$, of four final states are plotted as function of the variable coherent state argument $\alpha$. Two of the examples (a,b) have a high coherent state fidelity and naturally show a Q-function that resembles that of a coherent state. The other two (c,d) have a coherent state fidelity lower than 0.1. In these cases, we see that the main peaks are very similar to the coherent state peaks in the other figures. We also see, however, that there are areas in the complex plane away from the main peaks where the Q-functions are not negligible. The population in these regions shift the position and momentum expectation values so that $Q$-function may effectively become very small at these mean values. The maximum value of the $Q$-function, $max_\alpha\langle\alpha|\rho|\alpha\rangle$, over coherent state arguments would thus constitute a more fair measure of the coherent state character of the states obtained.

Our simulations indicate that the measurement back-action generates states close to coherent states. To understand how this can result from measurements of the position only, it is important to recall, that information about the position at one moment is converted into information about the momentum at later times due to the dynamics of the system. This is in competition with the adverse effect that position measurements at the same time cause an uncertainty in the momentum. Our measurements via the Cooper-pair box do not, however, cause an instant projection of the oscillator on position eigenstates, but rather constitute weak position measurements and we thus gradually obtain information about both position and momentum of the oscillator.

%
%
%The position of the oscillator is not a constant of motion. However, for interaction times $\tau$ such that $\omega\tau\ll1$ we can first consider the interaction between the Cooper-pair box and the cantilever mode and then consider the dynamics within the cantilever mode, that is we approximate the time-evolution operator with
%\begin{equation}
%U(\tau)=\exp(-i\omega\aup\adown\tau)\exp(-\frac{i\tau}{\hbar}(E_{c}\delta n\sigma_{z}+\lambda(\hat{a}^{\dag}+\hat{a})\sigma_{z}))
%\end{equation}
%Then the relative phase, $\Delta \phi$, only depends on the position of the cantilevermode. This way the measurment procedure will give information about the position of the oscillator.
%\\
%\\
%For a classical oscillator getting
%\begin{figure}[htb]
%	\begin{center}
%		\resizebox{!}{50mm}{\includegraphics{newfid.eps}}
%	\end{center}
%	\caption{This figure shows a histogram over the coherent state fidelity after 150 measurements for 500 simulations.}
%	\label{newfid}
%\end{figure}
To secure a symmetric probing of both position and momentum, we have simulated an interaction and detection protocol incorporating a wait time of $0.25/\omega$ between all measurement. This wait time ensures an alternating measurement sequence between the oscillator quadratures corresponding to equivalent probing of position and momentum properties.  In Fig. \ref{cohfid} a histogram of 500 attempts with this wait time between all meauserements is shown. For all 500 attempts the coherent state fidelity, defined via the state with the same mean position and momentum, is now above 0.91, as indicated in Fig. \ref{cohfid}.
\\
\\
\begin{figure}
	\begin{center}
		\resizebox{!}{70mm}{\includegraphics{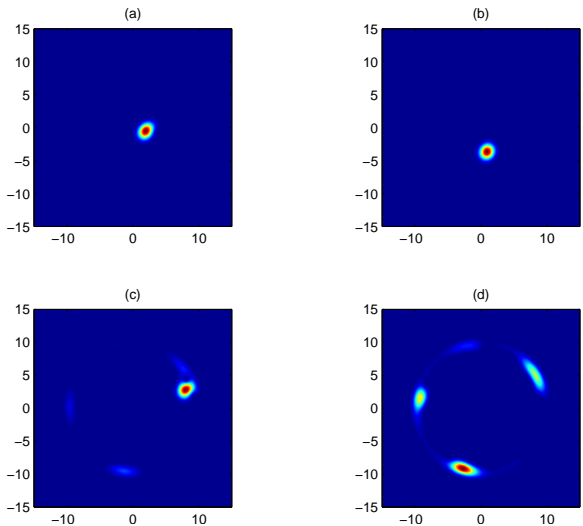}}
	\end{center}
	\caption{(Color online) Q-functions for four different simulations. In (a) and (b) the states are close to coherent states. The uncertainty products are 1.2 and 1.0, respectively, in units of $\hbar/2$ and the coherent state fidelities are	 0.96 and 0.99. In (c) and (d) the uncertainty products are 178 and 109, respectively, in units of $\hbar/2$, and we have state fidelities below 0.1 of coherent states with the same mean position and momentum.}
	\label{Qcanti}
\end{figure}

\begin{figure}[htb]
	\begin{center}
		\resizebox{!}{50mm}{\includegraphics{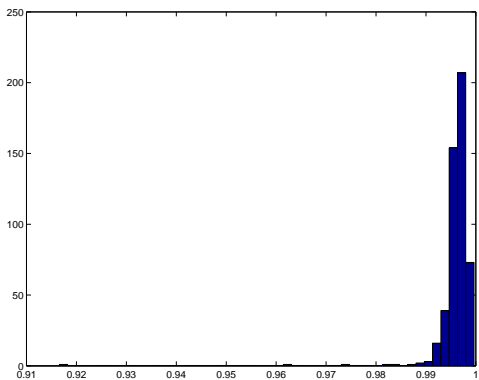}}
	\end{center}
	\caption{(Color online) Histogram of the coherent state fidelities for 500 simulations after 150 measurements have taken place. A wait time of $0.25/\omega$ is used between measurements.}
	\label{cohfid}
\end{figure}
In fact, not only do these wait times give a better result, they also represent a more realistic model of the measurement procedure allowing a finite read-out time for the Cooper-pair box.

\section{Feedback cooling to oscillator ground state}

Since we produce coherent states of the oscillator by our measurement back action, and the measurement record is known to the experimentalist, it is possible to convert the final state into the cantilever motional ground-state by application of a force. Since, however, our knowledge may not be fully accurate, and the force may be imprecise, it may be better to apply a feedback after each measurement and to adaptively let the dynamics produce and simultaneously verify the oscillator ground state. The feedback force can for example be accomodated by a piezo-resistive mount for the cantilever or by light pressure from a laser beam. It will be represented in the Hamiltonian by a term $\eta x$. Since the charge state of the Cooper-pair box is known when the feedback is applied the interaction with the Cooper-pair box can be readily compensated in the feedback term. If the expectation values of the position and momentum after a measurement procedure are estimated to be $\braket{x(0)}$ and $\braket{p(0)}$, the strength $\eta$ and the duration $\tau$ of the feedback force are given by,
\begin{equation}
\eta=-\frac{\braket{p(0)}^{2}+m^{2}\omega^{2}\braket{x(0)}^{2}}{2m\omega\braket{x(0)}}
\end{equation}
\begin{equation}
\tan(\omega\tau)=\frac{2m\omega\braket{x(0)}\braket{p(0)}}{m^{2}\omega^{2}\braket{x(0)}^{2}-\braket{p(0)}^{2}}
\end{equation}
where $m$ is the effective mass of the oscillator.

\begin{figure}
	\begin{center}
		\resizebox{!}{50mm}{\includegraphics{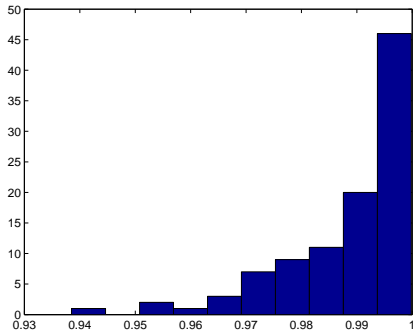}}
	\end{center}
	\caption{(Color online) The ground-state fidelity after 160 measurement and feedback operations. The simulations were performed with the same parameters as the ones shown in Fig. \ref{entropy}}	
	\label{histfeed}
\end{figure}

We have simulated this evolution, and in Fig. \ref{histfeed} a histogram of the ground-state population for a hundred iterations is shown. 160 measurements were performed in each iteration, and the same system parameters were used as for the other simulations. We see that these simulations results in a very high ground state population.

\subsection{The environment}

In the previous sections we treated the oscillator and Copper-pair box as if they were perfectly isolated from the environment between the measurement processes. The cantilever, however, is mechanically mounted, and it thermalizes to its environment at a rate determined by material and design parameters. The environment will thus limit our cooling as it tries to restore thermal equilibrium. The coherence time of the Cooper-pair box can be made as long as a few $\mu$s \cite{lifec}. Since the Cooper-pair box is put into a pure state each time its charge state is measured and the time between subsequent measurements is much smaller than the coherence time, the effect of the environments effect on the Cooper-pair box can be neglected.  Therefore we  concentrate on environmental effects on the cantilever mode, which we model by damping terms in a master equation for the oscillator density matrix. If $H$ is the system Hamiltonian, the appropriate master equation describing coupling with rate $\gamma$ to a black-body reservoir with a mean excitation of $\bar{n}$ quanta at the cantilever resonance frequency, is given by \cite{aassime2001rfs}:
\begin{eqnarray}
\frac{\partial\rho}{\partial t}=\frac{-i}{\hbar}[H,\rho]+\frac{\gamma}{2}(\bar{n}+1)(2\adown\rho\aup-\aup\adown\rho-\rho\aup\adown)\nonumber \\
+\frac{\gamma}{2}\bar{n}(2\aup\rho\adown-\adown\aup\rho-\rho\adown\aup)
\label{eq:master2}
\end{eqnarray}
From this equation we expect that both the thermal occupation of the reservoir $\bar{n}$ and the coupling constant $\gamma$ will play a role in the dynamics of the oscillator. 
We make the assumption that after a 100 measurements the system has reached steady state. Note that this is not a stationary state of the system, since each measurement and feedback process may both cause net excitation and de-excitation of the oscillator. In steady state, the population of the oscillator eigenstates after each feedback will thus fluctuate around a certain "typical" mean value. We expect that the memory of the oscillator of these steady state fluctuations will be limited to a couple of rounds of measurements and feedback, and hence the mean energy in steady state is effectively sampled by a long time average.

\begin{figure}
	\begin{center}
		\resizebox{!}{50mm}{\includegraphics{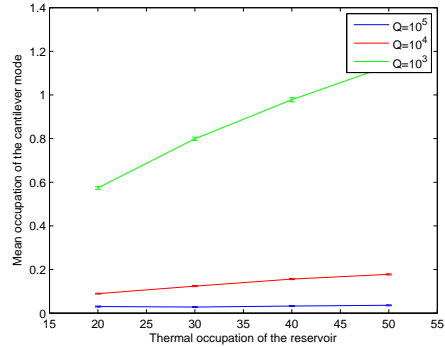}}
	\end{center}
	\caption{(Color online) Steady state excitation of the cantilever mode as a function of the thermal occupation of the reservoir for three different Q-values.}
	\label{Fig.61}
\end{figure}

Instead of the coupling strength it is customary to specify the Q-value for oscillators. This is related to the coupling constant by $Q=\omega/2\gamma$. Micromechanical oscillators have typical Q-values between $10^3$ and $10^5$. In Fig. \ref{Fig.61} we show the mean steady state energy of the oscillator plotted against the thermal occupation of the reservoir for three different Q-values.
Fig. \ref{Fig.61} provides a measure of how well the cooling counters the environmental heating. For a Q-factor of $10^{5}$ we get something very close to ground-state cooling. The mean excitation $\langle n \rangle$ is almost zero so that the occupation of the ground state is almost unity and the results show that the environment does not influence the system.
For $Q=10^4$ we obtain a mean excitation well below unity, while for $Q=10^3$ the mean occupation is on the order of unity but still far below $\bar{n}$ of the reservoir.

\section{Conclusion}

In summary, we have proposed to extract energy from a micromechanical oscillator by read-out of its position by means of a Cooper-pair box, and by acting back on the state of the oscillator according to the read-out. Unlike cooling schemes which explicitly damp the cantilever by the coupling to the readout device, our basic interaction does not have any such average cooling effect. The mechanism in fact extracts entropy rather than energy, in the sense that the interaction and measurements extract information about the cantilever and hence reduces its entropy until we reach a pure quantum state. By suitable timing of the measurements, this pure state happens to be very close to a coherent state, whose mean position and momentum are revealed by the detection record, and which can hence be deterministically displaced to the ground state - either at the end of the measurement record or by small steps after each measurement event. In comparison to the cooling schemes that work by a genuine damping mechanism, our cooling may work for a wider range of frequency and coupling parameters, and even for parameters where the average back action effect causes heating. Provided, of course, that we can provide the feedback force on the cantilever. In \cite{armour2002ead} a scheme is presented where similar physical measurements can be used to generate non-classical states of an oscillator which is initially in a coherent state. Our analysis uses the same architecture, and hence we provide a means to obtain the initial state needed for that and other protocols for generation of non-classical states.

\end{document}